# Hopf bifurcation analysis in a dual model of Internet congestion control algorithm with communication delay

Dawei Ding[a,b,*], Jie Zhu [a], Xiaoshu Luo [c], Yuliang Liu[a]

a. Department of Electronic engineering, Shanghai Jiao Tong University, Shanghai 200240, China

b. School of Electronic science and technology, Anhui University, Hefei 230039, China

c. Department of Physics and Electronic Science, Guangxi Normal University, Guilin 541004, China

**Abstract**

This paper focuses on the delay induced Hopf bifurcation in a dual model of Internet congestion control algorithms which can be modeled as a time-delay system described by a one-order delay differential equation (DDE). By choosing communication delay as the bifurcation parameter, we demonstrate that the system loses its stability and a Hopf bifurcation occurs when communication delay passes through a critical value. Moreover, the bifurcating periodic solution of system is calculated by means of perturbation methods. Discussion of stability of the periodic solutions involves the computation of Floquet exponents by considering the corresponding Poincaré-Lindstedt series expansion. Finally, numerical simulations for verify the theoretical analysis are provided.

1. Introduction

Congestion control algorithms and active queue management (AQM) for the Internet have been the focus of intense research since the seminal work of Kelly [1]. The primal and dual algorithms they have introduced are based on user utility and link pricing (explicit feedback) functions, where the sum of user utilities are maximized within the capacity (bandwidth) constraints of the links [2]. Congestion control schemes can be divided into three classes: primal algorithms, dual algorithms and primal-dual algorithms [3]. In primal algorithms, the users adapt the source rates dynamically based on the route prices (the congestion signal generated by the link), and the links select a static law to determine the link prices directly from the arrival rates at the links. While in dual algorithms, the links adapt the link prices dynamically based on the link rates, and the users select a static law to determine the source rates directly from the route prices and the source parameters. Primal-dual algorithms combine these two schemes and dynamically compute both user rates and link prices.

[*] Corresponding author. Tel. +86 21 34205433 Fax. +86 21 34205432
 Email: dwding@sjtu.edu.cn, dwdingsjtu@yahoo.com.cn

During the past decade, the stability of Internet congestion control system has drawn much attention from researchers. In [4], authors have analyzed the local stability of system where the end user implements the primal algorithm. Later Massoulie has extended these local stability results to general network topologies and heterogeneous delays [5]. In [6], authors have extended the framework of [1] and given a condition for its local stability under delay using the generalized Nyquist criterion. Sichitiu and Bauer have studied the asymptotic stability of congestion control systems with multiple sources [7]. Ranjan et.al have analyzed Kelly's optimization framework for a rate allocation problem and provide stability conditions with arbitrary fixed communication delays [8]. In [9], Wang and Eun have studied the local and global stability of TCP-newReno/RED under many flows.

Moreover, it is also of interest to find what will happen when the congestion control system loses its stability. Many research studies have been carried out to describe the complex dynamics of congestion control system, such as periodic oscillatory behaviors, bifurcation and chaos. One of the most cited paper about the chaotic nature of TCP is [10]. Using some discrete-time models, researchers have shown that TCP/RED systems become chaotic dynamics with variability in RED parameters [11-14]. As is well known, the entire congestion control system can also be regarded as a time-delay feedback control system, which can be described mathematically by delay differential equations (DDEs). Therefore, still some papers have focused on the Hopf bifurcation of the Internet congestion control system described by DDEs. In [15-17], the authors have shown that the existence of Hopf bifurcation in one-order and two-order REM with a single link and single source. Guo et al. have investigated the Hopf bifurcation of the exponential RED algorithm with communication delay [18, 19]. Raina has studied the local bifurcation of the fair dual with proportional and TCP fairness, and the delay dual algorithms by choosing a non-dimensional parameter as bifurcation parameter [20].

However, in [15-20], almost all authors applied the normal form theory and the center manifold theorem introduced in [21] to determinate the bifurcating periodic solutions and the stability and directions of the Hopf bifurcation. Recently, perturbation methods have shown the efficacy in studying the local dynamics near a Hopf bifurcation [22-23]. Compared with the center manifold reduction, the perturbation methods involve easy computations only, but yield the results of high accuracy [24]. Thus, in this paper we attempt to use perturbation methods to study the dynamic behavior in a fair dual algorithm of congestion control system. It should be noted that unlike the research in [20] in which the author has introduced a non-dimensional parameter, here we consider the communication delay as the bifurcation parameter, as the delay between the user and the source may vary depending on the network congestion status. Moreover, we still note that Wang et.al [17] have studied the Hopf bifurcation in a one-order REM congestion control algorithm model using perturbation method, but in their model they only considered the delayed term. While in our model described below, it contains not only the delayed term, but also the instantaneous term. Therefore, our study is more general.

The rest of this paper is organized as follows. In Section 2, by analyzing the corresponding characteristic equation of linearized equation of the dual model of the congestion control system, we derive the linear stability criteria and the existence of the Hopf bifurcation. The properties of bifurcating periodic solutions are determinated by using the perturbation methods in the following two sections. Numerical simulations and conclusions are given in Section 5 and 6, respectively.

## 2. Hopf bifurcation analysis

In this section, we consider the following dynamical representation of a fair dual congestion control algorithm [20]:

$$\frac{d}{dt} p(t) = kp(t)(x(t-\tau) - c) \tag{1}$$

where $x(t) = f(p(t))$ is a nonnegative continuous, strictly decreasing demand function and has at least third-order continuous derivatives. The scalar $c$ is the capacity of the bottleneck link and the variable $p$ is the price at the link. $k$ is a gain parameter.

Let $p^*$ be an non-zero equilibrium point of (1). Then it satisfies

$$x(p^*) = c$$

Expanding the equation (1) into first, second, third and higher-order terms about $p^*$, and defining $u(t) = p(t) - p^*$, we have

$$\dot{u}(t) = Lu(t) + H(u(t)) + h.o.t., \qquad (2)$$

where

$$\begin{cases} Lu(t) = b_1 u(t) + b_2 u(t-\tau) \\ H(u(t)) = b_3 u^2(t) + b_4 u(t) u(t-\tau) + b_5 u^2(t-\tau) \\ \qquad + b_6 u^3(t) + b_7 u^2(t) u(t-\tau) + b_8 u(t) u^2(t-\tau) + b_9 u^3(t-\tau) \end{cases}$$

with coefficients

$$b_1 = b_3 = b_6 = b_7 = 0,$$

$$b_2 = kp^* x'(p^*), b_4 = \frac{1}{2} kx'(p^*), b_5 = \frac{1}{2} kp^* x''(p^*),$$

$$b_8 = \frac{1}{6} kx''(p^*), b_9 = \frac{1}{6} kp^* x'''(p^*)$$

Therefore, the equation (2) can be rewritten as

$$\dot{u}(t) = b_2 u(t-\tau) + b_4 u(t) u(t-\tau) + b_5 u^2(t-\tau) + b_8 u(t) u^2(t-\tau) + b_9 u^3(t-\tau) + h.o.t. \quad (3)$$

Consider the linearized equation of (3), namely

$$\dot{u}(t) = b_2 u(t-\tau) \qquad (4)$$

whose corresponding characteristic equation is

$$\lambda - b_2 e^{-\lambda \tau} = 0$$

We begin with examining when this equation has pure imaginary roots $\lambda = \pm i\omega, \omega > 0$. Inserting them into the characteristic equation and separating the real and imaginary parts

$$\begin{cases} b_2 \cos(\omega \tau) = 0 \\ \omega + b_2 \sin(\omega \tau) = 0 \end{cases}$$

Therefore

$$\omega \tau = \frac{(2n+1)\pi}{2}, \ n = 0, 1, 2, \cdots$$

and

$$\omega + b_2 (-1)^n = 0$$

Thus yielding the critical value of $\tau$ as

$$\tau_c(n) = -\frac{(2n+1)\pi}{2b_2}, \quad n = 0, 2, 4, \cdots$$

Then we determinate whether the characteristic equation has roots with positive real part at $\tau_c(n)$. Let $\alpha + i\omega$ be a root of equation (4) with $\alpha$ and $\omega$ positive, then

$$\begin{cases} \alpha - b_2 e^{-\alpha\tau} \cos(\omega\tau) = 0 \\ \omega + b_2 e^{-\alpha\tau} \sin(\omega\tau) = 0 \end{cases} \quad (5)$$

From the first equation of (5), we know

$$\frac{(2n+1)\pi}{2} < \omega\tau < \frac{(2n+3)\pi}{2}, \quad n = 0, 2, 4, \cdots \quad (6)$$

And from the second equation of (5), we know

$$\omega\tau < \frac{(2n+1)\pi}{2} \quad (7)$$

Therefore, equation (4) may have roots with positive real part except for $n = 0$, i.e.,

$$\tau_0 = -\frac{\pi}{2b_2} \quad (8)$$

and in this case $\omega_0 = -b_2$.

However, we still have to determinate that for $\tau < \tau_0$ whether the system has roots with positive real part. Hence if $\alpha + i\omega$ is such a root with $\alpha > 0, \omega \geq 0$, noting that $b_2 < 0$, from equation (5) we get

$$\omega\tau = -\tau b_2 e^{-\alpha\tau} \sin(\omega\tau) < -b_2 \tau_0 = \frac{\pi}{2}$$

Thus $0 \leq \omega\tau < \pi/2$. Then from the first equation of (5), we get

$$\alpha = b_2 e^{-\alpha\tau} \cos(\omega\tau) < 0$$

which contradicts to $\alpha > 0$ as assumed. Therefore, equation (4) has no roots with positive real part when $\tau < \tau_0$. As $\tau_0$ is the smallest value of $\tau$ such that $\alpha = 0$, we get for $\tau < \tau_0$, all roots of equation (4) have negative real parts, which indicates that the equilibrium $p^*$ of system (1) is asymptotically stable whenever $\tau < \tau_0$.

Finally, we now satisfy the transversality condition of the Hopf spectrum, i.e.,

$$\text{Re}\left(\frac{d\lambda}{d\tau}\right)_{\tau=\tau_0} \neq 0$$

Therefore, evaluating

$$\frac{d\lambda}{d\tau} = -\frac{b_2\lambda e^{-\lambda\tau}}{1+b_2\tau e^{-\lambda\tau}}$$

Let $\lambda = \alpha + i\omega$, thus we have

$$\frac{d\lambda}{d\tau} = -\frac{b_2(\alpha+i\omega)e^{-\alpha\tau}(\cos(\omega\tau)-i\sin(\omega\tau))}{1+b_2\tau e^{-\alpha\tau}(\cos(\omega\tau)-i\sin(\omega\tau))}$$

From this we can obtain

$$\text{Re}\left(\frac{d\lambda}{d\tau}\right) = -\frac{b_2 e^{-\alpha\tau}(\alpha\cos(\omega\tau)+\omega\sin(\omega\tau)+b_2\tau\alpha e^{-\alpha\tau})}{(1+b_2\tau e^{-\alpha\tau}\cos(\omega\tau))^2+(b_2\tau e^{-\alpha\tau}\sin(\omega\tau))^2}$$

As we know, when $\tau = \tau_0$, $\alpha = 0$ and $\omega_0\tau_0 = \pi/2$. Therefore we obtain

$$\text{Re}\left(\frac{d\lambda}{d\tau}\right)_{\tau=\tau_0} = -\frac{b_2\omega_0}{1+(\omega_0\tau_0)^2} = \frac{b_2^2}{1+\pi^2/4} > 0$$

Based on above analysis, we can get the following theorem by applying the Hopf bifurcation theorem for delay differential equations [21].

**Theorem 1.** When the communication delay $\tau$ is smaller than the critical value $\tau_0 = -\pi/(2b_2)$, the equilibrium point $p^*$ of system (1) is asymptotically stable. When the delay $\tau$ passes through $\tau_0$, there is a Hopf bifurcation of system (1) at its equilibrium point $p^*$.

3. **Delay induced bifurcating periodic solutions**

In this section, based on the perturbation method, we analyze the bifurcating periodic solution of system (1) whose periodic depends on communication delay. Similar to the procedure in [25], we first rescale the variable $t$ by setting $s = \omega(\varepsilon)t$, where $\varepsilon$ is a small positive number so that solutions which are $2\pi/\omega$ periodic in $t$ will correspond to solutions which are $2\pi$ periodic in $s$. Therefore, equation (1) can be rewritten as

$$\omega\frac{du(s)}{ds} = k(u(s)+p^*)\left[x(u(s-\omega\tau)+p^*)-c\right] \tag{9}$$

And equation (3) can also be rewritten as

$$\omega \frac{du(s)}{ds} = b_2 u(t-\tau) + b_4 u(t)u(t-\tau) + b_5 u^2(t-\tau) \tag{10}$$
$$+ b_8 u(t)u^2(t-\tau) + b_9 u^3(t-\tau) + h.o.t.$$

The solution of equation (10) can be expressed in the form of perturbation series where

$$U(s,\varepsilon) = \varepsilon u_0(s) + \varepsilon^2 u_1(s) + \varepsilon^3 u_2(s) + \cdots = u(s) \tag{11}$$

The periodic solutions of nonlinear system (10) have periods depending on the parameter $\tau$; hence we perturb both the frequency and delay as follows

$$\begin{cases} \omega = \omega(\varepsilon) = \omega_0 + \omega_1 \varepsilon + \omega_2 \varepsilon^2 + \cdots \\ \tau = \tau(\varepsilon) = \tau_0 + \tau_1 \varepsilon + \tau_2 \varepsilon^2 + \cdots \end{cases} \tag{12}$$

where $\omega_0 = -b_2$ and $\tau_0 = -\pi/(2b_2)$.

From equation (11) and (12), we obtain

$$u(s - \omega\tau) = \varepsilon u_0(s - \omega\tau) + \varepsilon^2 u_1(s - \omega\tau) + \varepsilon^3 u_2(s - \omega\tau) + \cdots \tag{13}$$

where

$$u_i(s - \omega\tau) = u_i(s - \omega_0\tau_0) - u'_i(s - \omega_0\tau_0)\left[\varepsilon(\omega_1\tau_0 + \omega_0\tau_1) + \varepsilon^2(\omega_2\tau_0 + \omega_1\tau_1 + \omega_0\tau_2) + \cdots\right]$$
$$+ u''_i(s - \omega_0\tau_0)\left[\varepsilon(\omega_1\tau_0 + \omega_0\tau_1) + \cdots\right]^2 - \cdots$$

It is clear that all $u_i(s)$ are $2\pi$ periodic in the variable $s$. Substituting these expressions into (10) and using equation (11)-(13), we can obtain the following equations by equating the coefficients of the various terms involving powers of $\varepsilon$.

Therefore, it is obvious that $u_0(s)$ is governed by

$$\omega_0 \frac{du_0(s)}{ds} = b_2 u_0(s - \omega_0\tau_0) \tag{14}$$

In order to find a $2\pi$ periodic solution of (14), we let $u_0(s) = A_0 \sin s + B_0 \cos s$. Substituting it into equation (14), we obtain that $A_0$ and $B_0$ can be arbitrary. For the sake of simple calculation, we impose the initial condition $u_0(0) = 0$ and $u_0'(0) = 1$ to get that

$$u_0(s) = \sin s \tag{15}$$

Similarly, the term $u_1(s)$ in the perturbation series is governed by

$$\omega_1 \frac{du_0(s)}{ds} + \omega_0 \frac{du_1(s)}{ds} = -b_2 u'_0(s-\omega_0\tau_0)(\omega_1\tau_0 + \omega_0\tau_1) + b_2 u_1(s-\omega_0\tau_0) \quad (16)$$
$$+ b_4 u_0(s)u_0(s-\omega_0\tau_0) + b_5 u_0^2(s-\omega_0\tau_0)$$

Let $u_1(s) = A_1 \sin s + B_1 \cos s + C_1 \sin 2s + D_1 \cos 2s + E_1$. Then substitute it into (16) and with equation (15), we obtain

$$\omega_0(A_1 \cos s - B_1 \sin s) + 2\omega_0(C_1 \cos 2s + D_1 \sin 2s) = -\omega_1 \cos s - b_2 \sin s(\omega_1\tau_0 + \omega_0\tau_1)$$
$$+ b_2(-A_1 \cos s + B_1 \sin s - C_1 \sin 2s - D_1 \cos 2s + E_1) - \frac{b_4}{2}\sin 2s + \frac{b_5}{2} + \frac{b_5}{2}\cos 2s$$

where $\omega_0\tau_0 = \pi/2$. Comparing the corresponding coefficients, we get

$$\omega_1 = \tau_1 = 0, C_1 = -\frac{b_4 + 2b_5}{10b_2}, D_1 = \frac{-2b_4 + b_5}{10b_2}, E_1 = -\frac{b_5}{2b_2}.$$

Thus we pick the nontrivial solution

$$u_1(s) = A\sin_1 s + B_1 \cos s - \frac{b_4 + 2b_5}{10b_2}\sin 2s + \frac{-2b_4 + b_5}{10b_2}\cos 2s - \frac{b_5}{2b_2} \quad (17)$$

with $A_1$ and $B_1$ arbitrary.

The next system of equation can be obtained by comparing the coefficients of $\omega^3$ in equation (10), which gives

$$\omega_0 \frac{du_2(s)}{ds} + \omega_2 \frac{du_0(s)}{ds} = -b_2 u'_0(s-\omega_0\tau_0)(\omega_2\tau_0 + \omega_0\tau_2) + b_2 u_2(s-\omega_0\tau_0)$$
$$+ b_4 u_1(s)u_0(s-\omega_0\tau_0) + b_4 u_0(s)u_1(s-\omega_0\tau_0) + 2b_5 u_0(s-\omega_0\tau_0)u_1(s-\omega_0\tau_0) \quad (18)$$
$$+ b_8 u_0(s)u_0^2(s-\omega_0\tau_0) + b_9 u_0^3(s-\omega_0\tau_0)$$

Let $u_2(s) = A_2 \sin s + B_2 \cos s + C_2 \sin 2s + D_2 \cos 2s + E_2 \sin 3s + F_2 \cos 3s + G_2$ and with $\omega_0\tau_0 = \pi/2$, supplying it into equation (17) and comparing coefficients, then we obtain

$$\begin{cases} \omega_2 = -(b_4 + 2b_5)E_1 \\ -b_2(\omega_2\tau_0 + \omega_0\tau_2) + b_4 E_1 = 0 \end{cases}$$

Solving these equations, we then get that

$$\omega_2 = \frac{(b_4 + 2b_5)b_5}{2b_2}, \tau_2 = \frac{(2-\pi)b_4 b_5 - 2\pi b_5^2}{4b_2^3} \quad (19)$$

After finding the perturbed parameter values, we can write down the approximate solution of

equation (3) as

$$u(s) = \sqrt{\frac{\tau - \tau_0}{\tau_2}} u_0(s) + \left(\frac{\tau - \tau_0}{\tau_2}\right) u_1(s) + \cdots \qquad (20)$$

where $u_0(s)$ and $u_1(s)$ are given in equation (15) and (17) respectively and $\tau \approx \tau_0 + \varepsilon^2 \tau_2$.

As $\omega_1 = \tau_1 = 0$, we know that $\tau_2$ determinates the direction of the Hopf bifurcation and $\omega_2$ determinates the period of the bifurcating periodic solutions.

### 4. Stability of the bifurcating periodic solutions

In this section, we study the local stability of the bifurcating periodic solution by computing the Floquet exponents. We first rewrite the solution (20) into the following form

$$U(s,\varepsilon) = \varepsilon u_0(s) + \varepsilon^2 u_1(s) + \varepsilon^3 u_2(s) + \cdots = \varepsilon u(s)$$

In order to study the stability of the solution, we consider the corresponding Poincaré-Lindstedt series expansion. Let $x = \varepsilon u + Z$ be a solution of equation (9) with $Z$ a variation of the periodic solution $\varepsilon u$. Inserting $x$ in equation (9) yields

$$\varepsilon \frac{d(\varepsilon u + Z)}{ds} = k\left[\varepsilon u + Z + p^*\right]\left[x(\varepsilon u(s-\varepsilon\tau) + Z(s-\varepsilon\tau) + p^*) - c\right]$$

Since $\varepsilon u(s)$ is a solution of equation (9), we can simplify the above equation and obtain

$$\varepsilon \frac{d(Z(s))}{ds} = k\left[x(\varepsilon u(s-\varepsilon\tau) + p^*) - c\right]Z(s) \\ + k(\varepsilon u(s) + p^*)x'\left[\varepsilon u(s-\varepsilon\tau) + p^*\right]Z(s-\omega\tau) + \cdots$$

which can further be expanded to read

$$\varepsilon \frac{dZ}{ds} = [2b_4 \varepsilon u(s-\varepsilon\tau) + 3b_8 \varepsilon^2 u^2(s-\varepsilon\tau) + \cdots]Z(s) + [b_2 + 2b_4 \varepsilon u(s) \\ + 2b_5 \varepsilon u(s-\varepsilon\tau) + 3b_8 \varepsilon^2 u(s)u(s-\varepsilon\tau) + 3b_9 \varepsilon^2 u^2(s-\varepsilon\tau) + \cdots]Z(s-\omega\tau) + g(Z) \qquad (21)$$

where $g$ is a nonlinear term in $Z$ and is of $O(\varepsilon^3)$. If we ignore this term and let $Z = \hat{Z} + O(\varepsilon^3)$, then

$$\varepsilon \frac{d\hat{Z}}{ds} = [2b_4 \varepsilon u(s-\varepsilon\tau) + 3b_8 \varepsilon^2 u^2(s-\varepsilon\tau) + \cdots]\hat{Z}(s) + [b_2 + 2b_4 \varepsilon u(s) \\ + 2b_5 \varepsilon u(s-\varepsilon\tau) + 3b_8 \varepsilon^2 u(s)u(s-\varepsilon\tau) + 3b_9 \varepsilon^2 u^2(s-\varepsilon\tau) + \cdots]\hat{Z}(s-\omega\tau) \qquad (22)$$

We note that $\varepsilon u(s)$ is a solution of equation (9), i.e.,

$$\varepsilon\omega\frac{du}{ds} = k\left[\varepsilon u(s) + p^*\right]\left[x\left(\varepsilon u(s-\varepsilon\tau) + p^*\right) - c\right]$$

From this equation we have

$$\varepsilon\omega\frac{d}{ds}\left(\frac{du}{ds}\right) = \varepsilon k\left[x\left(\varepsilon u(s-\varepsilon\tau) + p^*\right) - c\right]\frac{du}{ds}$$
$$+ \varepsilon k(\varepsilon u(s) + p^*)x'\left[\varepsilon u(s-\varepsilon\tau) + p^*\right]\frac{du(s-\omega\tau)}{ds}$$

which can be rewritten to

$$\omega\frac{d}{ds}\left(\frac{du}{ds}\right) = [2b_4\varepsilon u(s-\varepsilon\tau) + 3b_8\varepsilon^2 u^2(s-\varepsilon\tau) + \cdots]\frac{du}{ds} + [b_2 + 2b_4\varepsilon u(s)$$
$$+ 2b_5\varepsilon u(s-\varepsilon\tau) + 3b_8\varepsilon^2 u(s)u(s-\varepsilon\tau) + 3b_9\varepsilon^2 u^2(s-\varepsilon\tau) + \cdots]\frac{du(s-\omega\tau)}{ds}$$

This shows that $du/ds$ is also a solution to equation (22). We will use this fact in that follows.

To study the linear stability of this periodic solution, we then compute the Floquet exponent. Let

$$\hat{Z} = e^{\eta s} q(s) \tag{23}$$

where $q(s)$ is $2\pi$ periodic in $s$. We call $\eta$ the Floquet exponent and $\exp(2\pi\eta)$ the Floquet multiplier. The stability of the trivial solution of equation (22) will depend on the sign of the real part of $\eta$.

We start with perturbing $\eta$ by putting

$$\eta = \omega\eta_1 + \omega^2\eta_2 + \cdots \tag{24}$$

Our task is to find an approximate value of $\eta$ by getting $\eta_1$ and $\eta_2$. To this end we look for a solution $q(s)$ in the form of

$$q(s) = h(\varepsilon)\frac{du}{ds} + q_0(s) + \varepsilon q_1(s) + \varepsilon^2 q_2(s) + \cdots$$
$$= h(\varepsilon)\frac{du}{ds} + \hat{q}(s) \tag{25}$$

where $q_i(s)$ is $2\pi$ periodic in $s$, and

$$\hat{q}(s) = q_0(s) + \varepsilon q_1(s) + \varepsilon^2 q_2(s) + \cdots$$
$$h(\varepsilon) = h_0 + \varepsilon h_1 + \varepsilon^2 h_2 + \cdots \tag{26}$$

Next, substituting the expression (23) in equation (22), we obtain

$$\omega \frac{dq(s)}{ds} + \omega \eta q(s) = [2b_4 \varepsilon u(s-\varepsilon\tau) + 3b_8 \varepsilon^2 u^2(s-\varepsilon\tau) + \cdots]q(s) + [b_2 + 2b_4 \varepsilon u(s) \\ + 2b_5 \varepsilon u(s-\varepsilon\tau) + 3b_8 \varepsilon^2 u(s)u(s-\varepsilon\tau) + 3b_9 \varepsilon^2 u^2(s-\varepsilon\tau) + \cdots]e^{-\eta\omega\tau}q(s-\omega\tau) \tag{27}$$

In addition, inserting the expression (25) in (27) and making use of the fact that $du/ds$ is a solution to equation (22) lead to the following equation in $\hat{q}(s)$

$$\omega \frac{d\hat{q}(s)}{ds} + \omega \eta \hat{q}(s) + \omega \eta h \frac{du}{ds} = [2b_4 \varepsilon u(s-\varepsilon\tau) + 3b_8 \varepsilon^2 u^2(s-\varepsilon\tau) + \cdots][\hat{q}(s) + h\frac{du}{ds}] \\
+ [b_2 + 2b_4 \varepsilon u(s) + 2b_5 \varepsilon u(s-\varepsilon\tau) + 3b_8 \varepsilon^2 u(s)u(s-\varepsilon\tau) + 3b_9 \varepsilon^2 u^2(s-\varepsilon\tau) + \cdots]e^{-\eta\omega\tau} \\
\left[\hat{q}(s-\varepsilon\tau) + h\frac{du(s-\varepsilon\tau)}{ds}\right] - h\left\{[2b_4 \varepsilon u(s-\varepsilon\tau) + 3b_8 \varepsilon^2 u^2(s-\varepsilon\tau) + \cdots]\frac{du}{ds} + [b_2 + \\
2b_4 \varepsilon u(s) + 2b_5 \varepsilon u(s-\varepsilon\tau) + 3b_8 \varepsilon^2 u(s)u(s-\varepsilon\tau) + 3b_9 \varepsilon^2 u^2(s-\varepsilon\tau) + \cdots]\frac{du(s-\varepsilon\tau)}{ds}\right\} \tag{28}$$

where $\hat{q}(s-\omega\tau) = q_0(s-\omega\tau) + \varepsilon q_1(s-\omega\tau) + \varepsilon^2 q_2(s-\omega\tau) + \cdots$

and for each $i$

$$q_i(s-\omega\tau) = q_i(s-\omega_0\tau_0) - q'_i(s-\omega_0\tau_0)\left[\varepsilon(\omega_1\tau_0 + \omega_0\tau_1) + \varepsilon^2(\omega_2\tau_0 + \omega_1\tau_1 + \omega_0\tau_2) + \cdots\right] \\
+ q''_i(s-\omega_0\tau_0)\left[\varepsilon(\omega_1\tau_0 + \omega_0\tau_1) + \cdots\right]^2 - \cdots$$

Finally we supply the following series with $\omega_1 = \tau_1 = 0$

$$e^{-\eta\omega\tau} = \exp\left\{-(\omega\eta_1 + \omega^2\eta_2 + \cdots)(\omega_0 + \omega_1\varepsilon + \omega_2\varepsilon^2 + \cdots)(\tau_0 + \tau_1\varepsilon + \tau_2\varepsilon^2 + \cdots)\right\} \\
= 1 - \varepsilon\eta_1\omega_0\tau_0 + \varepsilon^2\left(\frac{1}{2}\eta_1^2\omega_0^2\tau_0^2 - \eta_2\omega_0\tau_0\right) + O(\varepsilon^3)$$

as well as the series in equations (10), (11), (21), (24)-(26) in equation (28), and compare the coefficients of $\varepsilon^0$ and $\varepsilon^1$ to get the following equations respectively

$$\omega_0 \frac{dq_0(s)}{ds} = b_2 q_0(s - \omega_0\tau_0) \tag{29}$$

$$\omega_0 \frac{dq_1(s)}{ds} + \omega_0\eta_1 q_0(s) + \omega_0\eta_1 h_0 u'_0(s) = 2b_4 u_0(s-\omega_0\tau_0)q_0(s) \\
- b_2\eta_1\omega_0\tau_0[q_1(s-\omega_0\tau_0) + h_0 u'_0(s-\omega_0\tau_0)] + b_2 q_1(s-\omega_0\tau_0) \\
+ b_2 h_1 u'_0(s-\omega_0\tau_0) + [2b_4 u_0(s) + 2b_5 u_0(s-\omega_0\tau_0)]q_0(s-\omega_0\tau_0) \tag{30}$$

Observe that equation (29) is the same as equation (14) and hence a $2\pi$ periodic solution of

equation (29) is given by

$$q_0(s) = \sin s$$

Let $q_1(s) = A + B\cos s + C\sin s + D\cos 2s + E\sin 2s$ and with $\omega_0 \tau_0 = \pi/2$ we obtain

$$\eta_1 = 0$$

$$A = -\frac{b_5}{b_2}, \quad D = \frac{b_5 - 4b_4}{5b_2}, \quad E = -\frac{2b_4 + 2b_5}{5b_2}.$$

where $B$ and $C$ are arbitrary. We impose the initial condition $q_1(0) = 0$ and $q'_1(0) = 1$ to get

$$B = \frac{4b_4 + 4b_5}{5b_2}, \quad C = 1 + \frac{4b_4 + 4b_5}{5b_2}.$$

Therefore

$$q_1(s) = -\frac{b_5}{b_2} + \frac{4b_4 + 4b_5}{5b_2}\cos s + \left(1 + \frac{4b_4 + 4b_5}{5b_2}\right)\sin s + \frac{b_5 - 4b_4}{5b_2}\cos 2s - \frac{2b_4 + 2b_5}{5b_2}\sin 2s \quad (31)$$

To solve for $\eta_2$, we need to compare the coefficients of $\varepsilon^2$ in equation (28). With $\eta_1 = 0$, we obtain

$$\begin{aligned}
&\omega_0 \frac{dq_2(s)}{ds} + \omega_2 \frac{dq_0(s)}{ds} + \omega_0 \eta_2 q_0(s) + \omega_0 \eta_2 h_0 u'_0(s) = 2b_4 u_0(s - \omega_0 \tau_0) q_1(s) \\
&+ 3b_8 u_0^2(s - \omega_0 \tau_0) q_0(s) + b_2[q'_0(s - \omega_0 \tau_0)(\omega_2 \tau_0 + \omega_0 \tau_2) + q_2(s - \omega_0 \tau_0)] \\
&+ 2b_4 u_1(s) q_0(s - \omega_0 \tau_0) + 2b_5 u_0(s - \omega_0 \tau_0) q_1(s - \omega_0 \tau_0) \\
&+ 3b_8 u_0(s) u_0(s - \omega_0 \tau_0) q_0(s - \omega_0 \tau_0) + 3b_9 u_0^2(s - \omega_0 \tau_0) q_0(s - \omega_0 \tau_0)
\end{aligned} \quad (32)$$

Using equation (17) and (31), $q_0(s) = \sin s$, $\omega_0 \tau_0 = \pi/2$, and $u_0(s) = \sin s$, substituting

$$q_2(s) = A + B\cos s + C\sin s + D\cos 2s + E\sin 2s + G\cos 3s + H\sin 3s$$

into equation (32) and comparing the corresponding coefficients, we get

$$\eta_2 = \frac{5b_4 b_5}{2b_2^2} \quad (33)$$

Because $\eta_1 = 0$, it is clear that $\eta_2$ determinates the stability of the bifurcating periodic solutions. Recall that we have obtained values of $\tau_2$ and $\omega_2$ in the previous section, then we summarizing the above analysis in the following

**Theorem 2.** For the fair dual model of Internet congestion control system (1):

(1). $\tau_2$ determines the direction of the Hopf bifurcation. If $\tau_2 > 0$ ($\tau_2 < 0$) the Hopf bifurcation is supercritical (subcritical).

(2). $\eta_2$ determinates the stability of the bifurcating periodic solutions. If $\eta_2 < 0$ ($\eta_2 > 0$), the bifurcating periodic solutions are locally asymptotically stable (unstable).

(3). $\omega_2$ determines the period of the bifurcating periodic solution. If $\omega_2 < 0$ ($\omega_2 > 0$), the period of the solution increases (decreases).

5. **Numerical simulations**

This section gives some numerical simulations to verify the analysis above. We consider the fair dual which give a proportionally fair resource allocation, i.e.,

$$x(t) = \frac{1}{p(t)}$$

Let the link capacity is $1.25\ Mbps$ and the time unit is $40\ ms$. If the packet sizes are 1000 bytes each, then the link capacity can be expressed as $c = 50$ packets per time unit. In addition let the gain parameter $k = 0.01$.

The equilibrium point can be found by solving $1/p^* = c$, yielding $p^* = 0.02$. By direct calculation we can get

$$\tau_0 = 3.1416,\ \omega_0 = 0.5$$

and

$$\tau_2 = 7140.5,\ \eta_2 = -3125,\ \omega_2 = -937.5$$

Therefore, by theorem 1 the system equilibrium $p^*$ is asymptotically stable when $\tau < \tau_0$. Figure 1 and 2 are numerical simulations for this case with $\tau = 3$.

When $\tau$ passes through the critical value $\tau_0 = 3.1416$, the equilibrium loses its stability and a Hopf bifurcation occurs, i.e., a family of periodic solutions bifurcate from $p^*$. In addition, by theorem 2, the periodic orbits are stable since $\eta_2 < 0$. Also as $\tau_2 > 0$, the bifurcation at $p^*$ is supercritical and the bifurcating periodic solutions exist at least for $\tau$ slightly larger than the critical value. Figure 3 and 4 are numerical simulations with $\tau = 3.2$. They indicate that there is an orbitally stable limit cycle. Furthermore, since $\omega_2 < 0$, the period of the solutions increases as $\tau$ increases. For $\tau = 3.4$, the waveform plot and phase plot are shown in Figure 5 and 6, respectively. Comparing Figure 3 and Figure 5, we can see that the period of $\tau = 3.4$ is longer than that of $\tau = 3.2$.

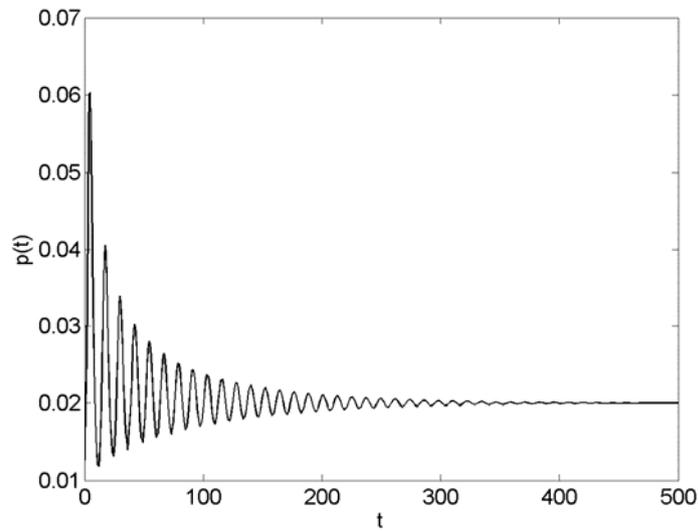

Figure 1 Waveform plot with $\tau = 3$

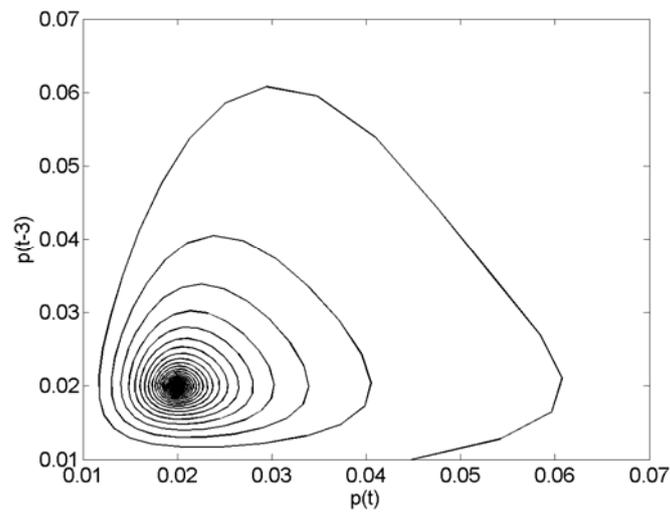

Figure 2 Phase plot with $\tau = 3$

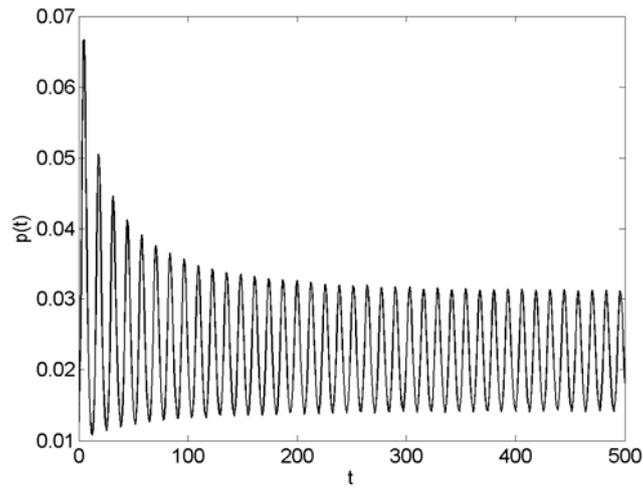

Figure 3 Waveform plot with $\tau = 3.2$

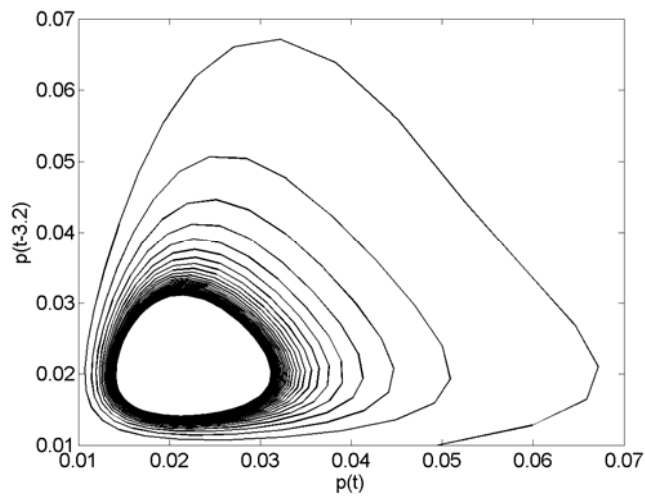

Figure 4 Phase plot with $\tau = 3.2$

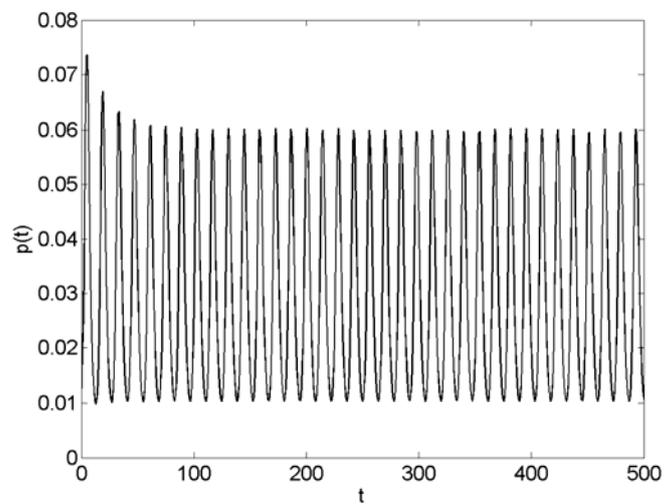

Figure 5 Waveform plot with $\tau = 3.4$

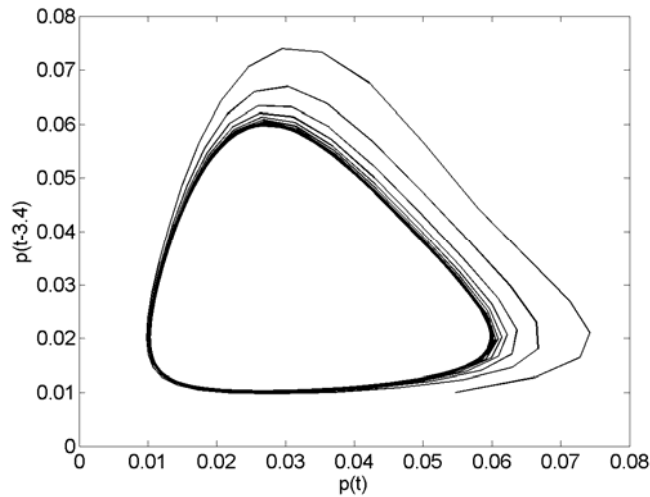

Figure 6 Phase plot with $\tau = 3.4$

## 6. Conclusions

In this paper, we focus our study on the Hopf bifurcation in a fair dual algorithm of Internet congestion control system. By using communication delay as a bifurcation parameter, we have shown that a Hopf bifurcation occurs in such a dual model of one order time-delay congestion control system when the communication delay of system passes through a critical value, i.e., a family of periodic orbits bifurcates from the equilibrium point. Furthermore, by means of perturbation method, we have analyzed the stability and direction of the bifurcating periodic solutions. Numerical simulations are accord with the theoretical results very well and verify the correctness of the analysis.

**Acknowledgments**

This work was supported by the National Natural Science Foundation of China with the grant numbers 70571017.